\documentclass[12pt]{article}

\usepackage{amsfonts,amssymb,amsmath}
\usepackage{graphicx}
\DeclareGraphicsExtensions{.pdf,.eps,.jpg}

\begin{document}

\title{\bf Engineering Quantum Photocells through Donor Multiplicity: Scaling Photocurrent and Power with N-Donor Architectures}
\author{Baharak Mohamad Jafari Navadel, Esfandyar Faizi, Baharam Ahansaz
\thanks{E-mail:b.ahansaz@azaruniv.ac.ir}
\\ {\small Physics Department, Azarbaijan Shahid Madani University, Tabriz, Iran}} \maketitle

\begin{abstract}
\noindent
Scaling quantum photovoltaic architectures beyond a few interacting molecular units remains a fundamental challenge for understanding how collective quantum dynamics translate into macroscopic electrical output. We present a scalable quantum photocell architecture consisting of $N$ identical donor molecules symmetrically arranged around a central acceptor, extending previous two- and three-donor models. Treating the system as a quantum heat engine and solving the Born-Markov master equation, we systematically compare the uncoupled and coupled regimes. Intermolecular coherent coupling produces delocalized collective excitonic states and modifies the excitation-transfer dynamics relative to independent donors. As the number of donors increases, both the steady-state photocurrent and output power increase sublinearly, with the coupled network consistently outperforming its uncoupled counterpart for every $N$ examined. The open-circuit voltage remains essentially unchanged, indicating that the power enhancement does not arise from an increase in the voltage output, but rather from modified excitation and charge-transfer dynamics. These results identify donor multiplicity, together with coherent intermolecular coupling, as complementary design parameters for engineering scalable quantum photovoltaic systems, and demonstrate that collective excitation dynamics can provide an intrinsic route to enhanced power generation beyond the additive contribution of independent molecular absorbers.

\end{abstract}

\newpage
\section{INTRODUCTION}
Photosynthesis is initiated by an ultrafast cascade of photophysical processes, wherein incident solar photons are efficiently captured and transduced into high-energy electrons. These charge carriers subsequently drive the molecular transformations underlying the later dark phases of the process. Likewise, photovoltaic cells (PV cells) convert solar radiation by photovoltaic effect analogous to classical heat engines. Optimizing emerging quantum solar cells at the nanoscale is crucial for maximizing light absorption and charge carrier dynamics, leading to significantly improved energy conversion efficiency. Precise nanoscale engineering enables better control over quantum effects, such as exciton generation and dissociation and coherence which are fundamental for enhancing performance beyond classical limits. By refining nanostructures, we can minimize energy losses and pave the way for next-generation sustainable energy technologies. Thermodynamic principles dictate the Shockley-Queisser limit as the fundamental efficiency ceiling for classical photovoltaics; however, by strategically harnessing quantum interference, it is possible to break the detailed balance constraint and exceed this theoretical boundary $\cite{Tokihiro}$. Long-lived quantum coherence, observed in photosynthetic systems under laser excitation, reveals nature’s ability to exploit quantum effects for efficient energy transfer $\cite{Calhoun,Abramavicius,Panitchayangkoon,Harel,Hayes,Romero}$. This quantum-driven mechanism enables the precise steering of excitonic pathways, minimizing energy losses and enhancing light-harvesting efficiency. Inspired by these phenomena, scientists seek to integrate quantum coherence into solar cell design, aiming to surpass classical efficiency limits and revolutionize photovoltaic technology $\cite{Mohseni,Plenio,Rebentrost,Zhu,Yeh}$. The Fenna-Matthews-Olson (FMO) complex, a pivotal pigment-protein system in photosynthetic bacteria, serves as a model for studying quantum coherence in energy conversion process. Its ability to sustain quantum superpositions and facilitate highly efficient exciton transport has aroused the design of photonically engineered photocells. The correlation between quantum effects and efficiency in nanostructures is paramount in shaping the future of quantum-based technologies. This synergy not only drives the development of highly efficient, sustainable energy systems but also paves the way for groundbreaking advancements in quantum information processing, potentially revolutionizing technologies across multiple domains $\cite{Fenna,Blankenship}$. Related concepts in quantum thermodynamics have also been explored in quantum batteries, where quantum correlations can influence energy transfer and storage performance $\cite{Shokri1,Shokri2}$.
A recent study by Dorfman et al. introduced a theoretical framework for enhancing light-induced processes in photocells modeled as quantum heat engines (QHEs) $\cite{Dorfman}$. The authors demonstrated that noise-induced quantum coherence specifically through Fano interference can significantly enhance charge separation efficiency in systems inspired by photosynthetic reaction centers (RCs). By linking population oscillations in photosynthetic complexes to improved light-harvesting performance, they reported an efficiency increase of at least $27\%$ over classical models. This quantum enhancement challenges the conventional thermodynamic limits, such as those established by Shockley and Queisser, and has been experimentally validated in optical platforms. In this respect, Scully et al. theoretically showed that quantum coherence can boost the efficiency of the quantum photocells $\cite{Scully1,Scully2,Svidzinsky}$. Building on Scully et al.'s findings, Creatore et al. introduced a biomimetic photocell model that leverages a delocalized dark quantum state, formed through dipole coupling between two donor molecules which was predicted to enhance the efficiency by approximately $35\%$ $\cite{Creatore}$. Next, Fruchtman et al. demonstrated that a photocell integrating an asymmetric chromophore pair can outperform designs based on symmetric dimers or independent molecules $\cite{Fruchtman}$. The interaction between coherent and incoherent energy transfer has been extensively studied in molecular crystals and aggregates, revealing that exciton delocalization is governed by the balance between exciton coupling and energetic disorder. As delocalization increases, coherent effects play a more prominent role in transport dynamics. The study by Zhang et al. $\cite{Zhang}$, which demonstrated that delocalized quantum states arising from donor-donor coupling can enhance the performance of a photocell with three coupled electron donor levels after initially considering two, raises a broader question concerning the role of donor multiplicity in quantum photovoltaic architectures. In particular, it remains unclear whether the performance enhancement associated with increasing the number of donor molecules persists as the donor ensemble is systematically enlarged, and, more importantly, whether such scaling is fundamentally different for uncoupled and coupled interacting donors.
Motivated by this idea, we extend this line of inquiry by generalizing the model to $N$ electron donors and systematically investigating how increasing the number of donor levels influences system performance and quantum efficiency. Particular attention is given to the role of intermolecular dipole-dipole coupling, which establishes an interaction pathway among the donor molecules and contributes to the emergence of collective dynamics. The resulting interactions enable the formation of delocalized excited states across the donor ensemble, providing a physical basis for collective quantum dynamics.
Within this multi-donor architecture, quantum coherence plays a central role in sustaining these collective excitations, allowing the interacting donors to respond in a correlated manner rather than as independent constituents. Within this framework, we perform numerical calculations to investigate the dynamics and performance of the resulting complex system comprising $N$ electron donors. These results indicate that increasing the number of coherently coupled donors can provide a viable route for exploiting collective quantum effects and optimizing the performance of multi-donor photocells.
Rather than focusing on a specific small donor configuration, we introduce a scalable multi-donor architecture that incorporates intermolecular dipole-dipole interactions, providing a systematic framework for examining how donor multiplicity and collective excitation dynamics influence energy transfer and power generation, and for elucidating the mechanisms underlying quantum-enhanced energy conversion.

\section{MODEL DESCRIPTION}
The realistic photocell model proposed here is a QHE paradigm inspired by photosynthesis to investigate and potentially enhance the light-harvesting efficiency of quantum photocells. To elucidate the key difference between conventional models and our proposed approach, we examine a simple multi-donor quantum photovoltaic model. Previous studies on quantum-enhanced photocells have primarily focused on systems with a limited number of donor molecules, typically two. In contrast, our work extends the cyclic engine model to an arbitrary number of $N$ donor molecules which emulates the photosynthetic reaction center apparatus, aligning with the foundational framework established by Dorfman $\cite{Dorfman}$. Crystallographic analysis suggests that molecules within aggregates predominantly exhibit collective alignment rather than independent orientation, thereby enhancing optoelectronic properties through tunable light-matter interactions $\cite{Friend,Uoyama,Tessler}$. These changes can influence absorption, emission, energy transfer, and photophysical processes such as exciton dynamics.
Here, we designed an extended donor-acceptor system consisting of $N$ identical donor molecules with degenerate excitation energies, each selected based on complementary absorption spectra and optimized energy-level alignment with the chosen acceptor material. In the uncoupled configuration, the excitations are essentially localized and the donors behave as independent excitation centers. As the intermolecular separation is reduced, the spatial proximity of the donor molecules can give rise to appreciable intermolecular interactions. In particular, the transition dipole moments associated with optically allowed donor excitations can interact through long-range dipole-dipole coupling $\cite{Abramavicius,Navadel}$. This interaction gives rise to finite excitonic coupling between neighboring donor molecules, and consequently, the initially degenerate localized excited states can superpose and evolve into collective excitonic eigenstates that extend over multiple donor molecules. Therefore, the excitation is transformed from a localized molecular excitation into a collective excitation distributed over the coupled donor network. To investigate the consequences of this collective behavior, we now consider a well-defined donor-acceptor structure consisting of an array of $N$ identical donor molecules, each capable of absorbing photons and transferring excited electrons to a common acceptor site as depicted in Figure 1. The optical excitation of collective diverse donor network induced by solar radiation is modeled using a two-level system, where the ground state is denoted as $\vert b\rangle$, and the excited states are represented as $\vert a_{1}\rangle$ and $\vert a_{2}\rangle$,..., $\vert a_{N}\rangle$. The exciton dynamics of the collective multi-donor structure is governed by,
\begin{eqnarray}\label{eq.0}
\mathcal{H}_{D}=\sum_{i=1}^{N}\hbar\omega_{i}\sigma_{i}^{+}\sigma_{i}^{-}+\sum_{i=1}^{N-1}J_{i,i+1}(\sigma_{i}^{+}\sigma_{i+1}^{-}+\sigma_{i}^{-}\sigma_{i+1}^{+}),
\end{eqnarray}
where $\omega_{i}$ represents the excitation energy of donor $i$ and $\sigma_{i}^{+}=(\sigma_{i}^{-})^\dag=\vert a_{i}\rangle \langle b \vert$ with $i=1,...,N$ are the raising (lowering) operators. Here, the donor-donor coupling is assumed to be uniform and constant for all neighboring donor pairs, such that $J_{i,i+1}=J$ for $i=1,...,N-1$. Depending on the strength of the intermolecular interaction, we consider two physically distinct regimes of the donor ensemble, corresponding to the absence and presence of intermolecular excitonic coupling. In the uncoupled regime, the donor-donor coupling is negligible, ($J=0$), and the donor molecules behave as independent excitation centers. In this limit, an optical excitation remains essentially localized on an individual donor, and the corresponding excited states $(\vert a_i\rangle)$ constitute approximately independent localized states. In contrast, in the coupled regime, finite intermolecular coupling, $(J\neq 0)$, arises from the long-range transition dipole-dipole interaction between neighboring donor molecules. This coupling changes the nature of the relevant quantum states from localized donor excitations to potentially delocalized collective excitations. The resulting collective excitonic states enable coherent excitation dynamics across the donor ensemble. Thus, the uncoupled and coupled regimes provide two cases for direct comparison between localized independent excitations and delocalized collective excited states, thereby clarifying the extent to which the enhancement of energy transfer and power generation originates from the mere presence of multiple donors or from genuinely collective quantum effects.

As depicted in Figure 1, the donor molecules $D_{1}$ through $D_{N}$ constitute an ensemble of identical, initially uncoupled molecules arranged symmetrically around a central acceptor molecule A. The excitation cycle is initiated by the absorption of concentrated solar photons, promoting the system from its ground state $\vert b \rangle$ to the donors excited states, denoted as $\vert a_{1} \rangle$, $\vert a_{2} \rangle$,..., $\vert a_{N} \rangle$ with energies $E_{1}$, $E_{2}$,..., $E_{N}$. Subsequent to photoexcitation, the excited electrons are transferred from the donors to the acceptor via electronic coupling mechanisms, accompanied by phonon emission, in accordance with the processes described in Ref. $\cite{Dorfman}$. Following the charge separation process, the excited electrons are harnessed to perform useful work, driving the transition of the system from the charge-separated state $\vert \alpha \rangle$ to a semi-stable state $\vert \beta \rangle$. Additionally, recombination between the acceptor and donor is accounted for, occurring at a decay rate of $\chi \Gamma$, where $\chi$ denotes a dimensionless fraction. Ultimately, the semi-stable state $\vert \beta \rangle$ decays via the intermediate channel $\gamma_C$, returning the system to the charge-neutral ground state $\vert b \rangle$, thereby completing the cycle.

In what follows, we develop a comprehensive kinetic model to describe the time-dependent evolution of the average occupation numbers associated with the specific quantum level structure under consideration. To make this problem analytically tractable, we adopt the well-established Born-Markov approximation framework. This approach is predicated on two key assumptions: first, that the interaction between the open quantum system (in this case, the model photocell) and its surrounding environment is sufficiently weak such that perturbative treatment is valid; and second, that the environment exhibits a very short correlation time relative to the system's dynamics, effectively allowing us to neglect any memory effects and treat the environment as Markovian. These approximations enable a coarse-grained description of the system's dynamics, wherein the evolution depends solely on the current state, and not on its past history. Under these conditions, we can derive a time-local master equation that governs the evolution of the system's reduced density matrix $\rho(t)$, encapsulating the influence of the environment through dissipative terms. The resulting equation forms the basis for analyzing the photocell's population dynamics and energy transfer processes $\cite{Breuer,Carmichael,Oh}$
\begin{eqnarray}\label{eq.1}
\dfrac{\partial\rho}{\partial t}=(\dfrac{i}{\hbar}) [\rho,\mathcal{H}] +\mathcal{L}^{h}(\rho)+\mathcal{L}^{c}(\rho)+\mathcal{L}^{D}(\rho).
\end{eqnarray}
The Hamiltonian governing the dynamics of the photocell system is expressed as follows:
\begin{eqnarray}\label{eq.8}
 \mathcal{H}=\mathcal{H}_{D}+\mathcal{H}_{A}
 \end{eqnarray}
where $\mathcal{H}_{A}=\hbar\omega_{\alpha}\sigma_{\alpha}^{+}\sigma_{\alpha}^{-}+\hbar\omega_{\beta}\sigma_{\beta}^{+}\sigma_{\beta}^{-}$ is the Hamiltonian of the accepter and $\sigma_{\alpha}^{+}=(\sigma_{\alpha}^{-})^\dag=\vert \alpha\rangle \langle b \vert$ and $\sigma_{\beta}^{+}=(\sigma_{\beta}^{-})^\dag=\vert \beta\rangle \langle b \vert$ are the related raising (lowering) operators. The Lindblad superoperator, denoted by $\mathcal{L}^{h}(\rho)$ characterizes the irreversible dynamics arising from the interaction between the quantum system and its surrounding thermal reservoir, specifically the hot bath. This operator accounts for the dissipative processes induced by the thermal fluctuations of the environment and governs the flow of energy between the system and the bath. Within the Born-Markov and secular approximations, $\mathcal{L}^{h}(\rho)$ takes the standard Lindblad form and is given by:
\begin{eqnarray}\label{eq.9}
\mathcal{L}^{h}(\rho)=\sum^{N}_{i=}(\dfrac{\gamma_{ih}}{2})(n_{ih}+1)\bigg[2\sigma_{i}^{-}\rho\sigma_{i}^{+}-\sigma_{i}^{+}\sigma_{i}^{-}\rho-\rho\sigma_{i}^{+}\sigma_{i}^{-}\bigg]  +(\dfrac{\gamma_{ih}}{2})n_{ih}\bigg[2\sigma_{i}^{+}\rho\sigma_{i}^{-}-\\ \nonumber
\sigma_{i}^{-}\sigma_{i}^{+}\rho-\rho\sigma_{i}^{-}\sigma_{i}^{+}\bigg],
\end{eqnarray}
where $\sigma_{i}^{+}=(\sigma^{-})^\dag=\vert a_{i}\rangle \langle b \vert$ is the corresponding raising (lowering) operator. Furthermore, the parameter $\gamma_{ih}$ denotes the transition rate between ground state $\vert b\rangle$ and excited states $\vert a _{i} \rangle$, induced by interaction with the hot thermal reservoir. The thermal population of the reservoir is characterized by the average photon number $n_{ih}$, which follows the Plank distribution at temperature $T_{h}$ $\cite{Breuer,Mandel}$. This interaction gives rise to thermal excitation processes governed by Bose-Einstein statistics. Similarly, the influence of the cold bath on the system dynamics is encapsulated by the Lindblad superoperator $\mathcal{L}_c(\rho)$, which accounts for dissipative transitions primarily relaxation from the excited states back to the ground state, as mediated by the colder thermal environment:
\begin{eqnarray}\label{eq.10}
\mathcal{L}^{c}(\rho)=\mathcal{L}_{1}^{c}(\rho)+\mathcal{L}_{2}^{c}(\rho),
\end{eqnarray}
here $\mathcal{L}_{1}^{c}(\rho)$ and  $\mathcal{L}_{1}^{c}(\rho)$ are given by,
\begin{eqnarray}\label{eq.11}
\mathcal{L}_{c}^{1}(\rho)=\sum^{N}_{i=}(\dfrac{\gamma_{ic}}{2})(n_{ic}+1)\bigg[2L_{i}^{-}\rho L_{i}^{+}-L_{i}^{+}L_{i}^{-} \rho-\rho L_{i}^{+}L_{i}^{-}\bigg]  +(\dfrac{\gamma_{ic}}{2})n_{ic} \bigg[2L_{i}^{+}\rho L_{i}^{-}-L_{i}^{-}L_{i}^{+} \rho-\\ \nonumber
\rho L_{i}^{-}L_{i}^{+}\bigg]
\end{eqnarray}
and
\begin{eqnarray}\label{eq.12}
\mathcal{L}_{c}^{2}(\rho)=(\dfrac{\Gamma_{c}}{2})(N_{c}+1)\bigg[2\tau^{-}\rho \tau^{+}-\tau^{+}\tau^{-} \rho-\rho \tau^{+}\tau^{-}\bigg]  +(\dfrac{\Gamma_{c}}{2})N_{c} \bigg[2\tau^{+}\rho \tau^{-}-\tau^{-}\tau^{+} \rho-\rho \tau^{-}\tau^{+}\bigg]
\end{eqnarray}
Here, $\gamma_{ic}$ and $\Gamma_{c}$ are spontaneous decay rates and $L_{i}^{+}=(L_{i}^{-})^\dag=\vert a_{i}\rangle \langle \alpha \vert$ and  $\tau^{+}=(\tau^{-})^\dag=\vert \beta\rangle\langle b \vert$  are the jump operators. Accordingly, $n_{ic}=n_{1c}, n_{2c},...,n_{Nc}$ and $N_{c}$ denote the average thermal occupation numbers at temperature $T_{c}$ corresponding to energy gaps $\Delta E=E_{a_{i}}-E_{\alpha}$ and $\Delta E=E_{\beta}-E_{b}$, respectively. The average thermal occupations associated with these energy differences are given by,
\begin{eqnarray}\label{eq.11}
n=\dfrac{1}{e^{\Delta E/K_{B}T_{c}}-1},
\end{eqnarray}
Here, $K_{B}$ stands for the Boltzmann constant which sets the scale for thermal energy. Consequently, the Lindblad dissipator $\mathcal{L}^{D}(\rho)$ can be explicitly expressed in the following detailed form:
\begin{eqnarray}\label{eq.12}
\mathcal{L}^{D}(\rho)=(\dfrac{\Gamma}{2})\bigg[2\lambda_{1}^{-}\rho\lambda_{1}^{+}-\lambda_{1}^{+}\lambda_{1}^{-}\rho-\rho\lambda_{1}^{+}\lambda_{1}^{-}\bigg] +(\dfrac{\chi\Gamma}{2})\bigg[2\lambda_{2}^{+}\rho \lambda_{2}^{-}-\lambda_{2}^{+}\lambda_{2}^{-}\rho-\rho \lambda_{2}^{+}\lambda_{2}^{-}\bigg].
\end{eqnarray}
In the equation under consideration, the operators  $\lambda_{1}^{+}=(\lambda_{1}^{-})^\dag=\vert \alpha\rangle \langle \beta \vert$ and $\lambda_{2}^{+}=(\lambda_{2}^{-})^\dag=\vert \alpha\rangle \langle b \vert$ function as quantum jump operators within the Lindblad formalism, mediating incoherent transitions between the corresponding energy eigenstates of the system. Specifically,  $\lambda_{1}^{+}$ facilitates transitions from state $\vert \beta\rangle$ to $\vert \alpha \rangle$ while, $\lambda_{2}^{+}$ governs transitions from $\vert b \rangle$ to $\vert \alpha \rangle$. The transition rate  $\Gamma$ quantitatively characterizes the dissipative coupling responsible for population transfer from the excited state $\vert \alpha \rangle$ to the lower-energy state $\vert \beta \rangle$ , typically induced by the interaction with an external reservoir or environment. Meanwhile, the parameter $\chi$ encapsulates the recombination dynamics of photogenerated charge carriers electrons and holes occurring at the donor-acceptor interface. This recombination process plays a crucial role in determining the overall performance of the photocell, as it directly impacts the charge separation efficiency and, consequently, the net photocurrent generation. The system's dynamics are governed by the Pauli master equation, which ensures the complete positivity and physical consistency of the population evolution. Furthermore, the proposed model operates within the framework of a quantum heat engine, simultaneously coupled to both a hot and a cold thermal reservoir. This configuration enables the extraction of useful work from thermal gradients via light-matter interactions, while maintaining thermodynamic consistency and detailed balance in the steady state.

\begin{figure}
\centering $\mathbf{(a)}$\\{
\includegraphics[width=3.4in]{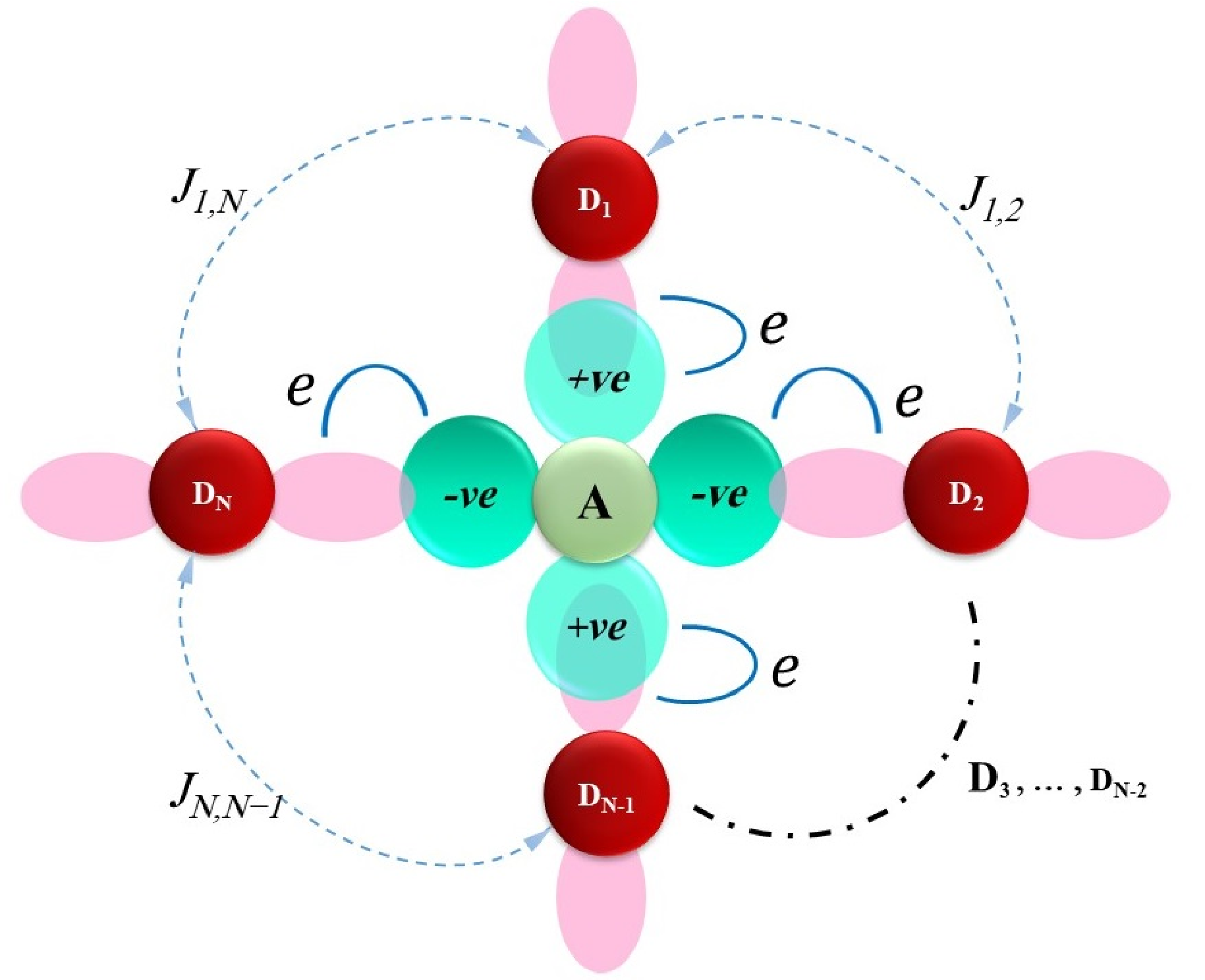}
\label{fig:first_sub}
}\\ \centering $\mathbf{(b)}$\\{
\includegraphics[width=3in]{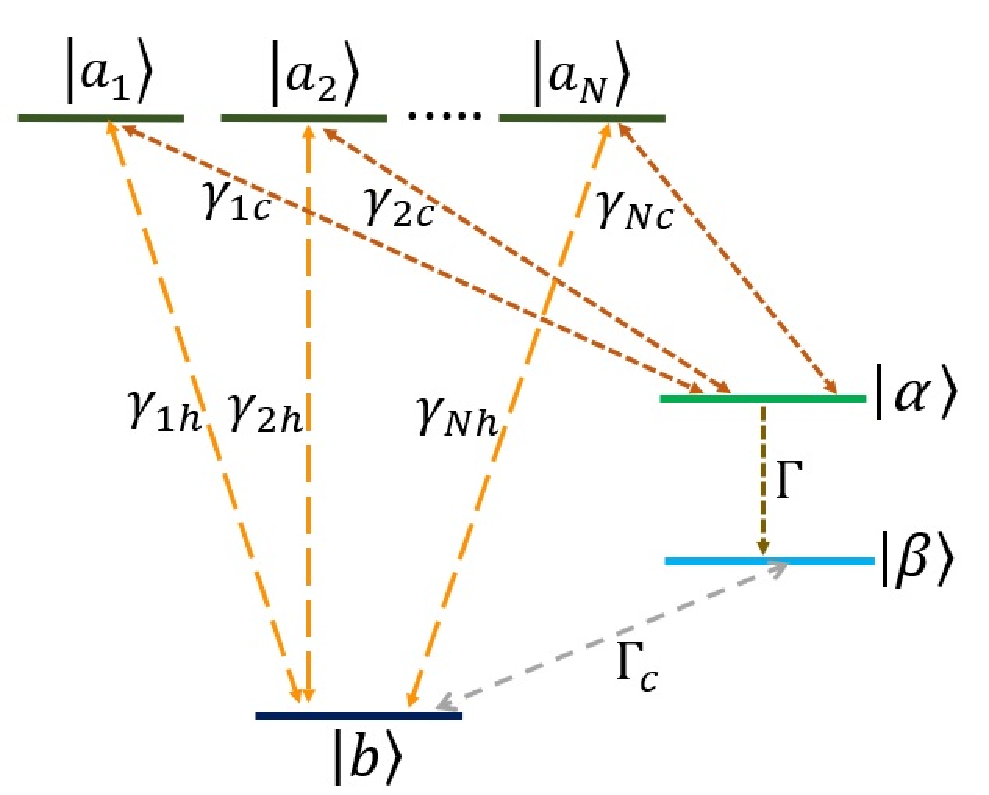}
\label{fig:first_sub}
}
\caption{\small \bf Schematic of the reaction center (a) $N$ independent and identical donor molecules surrounding the acceptor molecule facilitate electron transfer through cooperative action, where electrons are transferred from the donors to the acceptor. (b) The level architecture and electron paths in multi-donor based PV cell.}
\end{figure}

\section{RESULTS AND DISCUSSION}
Prior to undertaking the main analysis, it is essential to numerically solve the Pauli master equation Eq.(\ref{eq.1}) in order to capture both the transient dynamics and the steady-state properties of the system. As the evolution proceeds for a sufficiently long time, the system reaches a steady-state regime in which the populations approach time-independent values, indicating a balance between the different excitation, relaxation, recombination, and charge-extraction processes. Notably, this numerical solution is performed here for the first time, despite the inherent complexity and sophistication of the equation, marking a significant methodological advancement. The complete set of parameters employed in the simulations is detailed in Table 1.
\begin{center}
\begin{tabular}{ |l|l| }
  \hline
  \multicolumn{2}{|c|}{Table 1. The used parameters.} \\
  \hline
  $E_{a_{1},...,a_{N}} - E_{b}$ & 1.8 (eV) \\
  $E_{a_{1},...,a_{N}} - E_{\alpha}$ & 0.2 (eV) \\
  $E_{\beta}$ -$E_{b}$ & 0.2 (eV) \\
  $E_{\alpha}$-$E_{\beta}$ & 1.4 (eV)\\
  $\gamma_{1h},...,\gamma_{Nc}$ & $0.62\times 10^{-6}$ \\
  $\gamma_{1c},...,\gamma_{Nc}$ & $6\times 10^{-3}$ \\
  $\Gamma_{c}$ & 0.025 \\
  $\Gamma$ & 0.12\\
  $\chi$ & 0.2\\
  \hline
\end{tabular}
\end{center}
By introducing the concepts of photochemical current and photochemical voltage, we can effectively characterize the reaction center in terms of an equivalent electrical circuit. In this framework, the reaction center is assigned an effective current and voltage that correspond to its photochemical activity. The external load resistance is modeled through the electron decay rate, denoted by $\Gamma$, which quantifies the rate at which electrons transition from the cathode to the anode. The net electronic current associated with transitions from quantum state $\vert \alpha \rangle$ to $\vert \beta \rangle$ is given by:
\begin{eqnarray}\label{eq.12}
j =e \Gamma P_{\alpha}
\end{eqnarray}
Here, $e$ is the elementary charge (the fundamental charge of an electron), and $P_{\alpha}$ represents the steady-state probability of the system being found in state $\vert \alpha \rangle$. This probability is derived from the diagonal elements of the system's density matrix, such that for any state $i$ (where $i \in {a_{1}, a_{2},...,a_{N}, \alpha, \beta, b}$), the population is $P_{i}=\rho_{i,i}$ where $\rho_{i,i}$ denotes the $i^{th}$ diagonal element of the density matrix $\rho$ , which encapsulates the statistical properties of the system in steady state. It is important to note that the parameter $\Gamma$ spans a range from $\Gamma=0$, corresponding to the open-circuit limit, to large values of $\Gamma$, which characterize the short-circuit regime. The voltage generated across the solar cell is dictated by the difference in chemical potentials between the two electronic states associated with the external terminals. This potential difference can be formally expressed as: $eV\equiv e(V_{\beta}-V_{\alpha})$=$\mu_{\alpha}-\mu_{\beta}$, where $V_{\alpha}$ and $V_{\beta}$ represent the electrostatic potentials at states  $\vert \alpha \rangle$ and $\vert \beta \rangle$, respectively, and $\mu_{\alpha}$ and $\mu_{\beta}$ denote the corresponding chemical potentials. To establish a thermodynamic link between the voltage and the microscopic properties of the system, we consider that the populations of states  $\vert \alpha \rangle$ and $\vert \beta \rangle$ follow Boltzmann statistics at the temperature of the cold reservoir $T_{c}$. Under this assumption, $P_{\alpha}=e^{-(E_{\alpha}-\mu_{\alpha})/K_{B}T_{c}}$ and $P_{\beta}=e^{-(E_{\beta}-\mu_{\beta})/K_{B}T_{c}}$ are the occupation probabilities where $E_{\alpha}$ and $E_{\beta}$ are the respective energy levels of the two states. This formulation allows the photovoltage to be expressed directly in terms of the energy level spacing and the statistical populations of the electronic states, thereby linking the macroscopic electrical output to the underlying microscopic thermodynamic parameters $\cite{Scully2,Creatore}$
\begin{eqnarray}\label{eq.13}
eV=E_{\alpha}-E_{\beta}+K_{B}T_{c}\ln\dfrac{P_{\alpha}}{P_{\beta}}.
\end{eqnarray}

In the absence of sunlight, the quantum photocell evolves toward thermal equilibrium with its environment, predominantly characterized by the temperature of the phonon bath, denoted by $T_{c}$. Under these equilibrium conditions, no net photogeneration occurs, and the chemical potentials of the relevant electronic states equalize, resulting in a vanishing photovoltage $(V=0)$. This behavior reflects the system's return to detailed balance, where all transitions between states occur at equal and opposite rates. Therefore, the emergence of a non-zero voltage under illumination serves as a direct indicator of the system's deviation from thermodynamic equilibrium. Specifically, the voltage $V$ quantifies the degree of non-equilibrium induced by photoexcitation, relative to the reference thermal state defined by the cold bath temperature $T_{c}$. It effectively captures the thermodynamic driving force responsible for charge separation in the solar cell. Upon absorption of a photon, an electron is excited to a higher energy state and is rapidly transferred to an acceptor site, provided the transfer rate exceeds the competing relaxation rates. This ultrafast electron transfer facilitates charge separation before significant recombination can occur. As a result, the generated photocurrent becomes approximately proportional to the elementary charge $e$ multiplied by the electron generation rate. For typical quantum photocells, this photocurrent lies in the range of microamperes. To accurately determine the steady-state current and voltage, one must solve the quantum master equation governing the system's dynamics. The master equation provides the time evolution of the system's density matrix $\rho(t)$. In this regime, the populations of the relevant energy eigenstates specifically, $\rho_{\alpha,\alpha}$ and $\rho_{\beta,\beta}$ are extracted from the diagonal elements of the steady-state density matrix. These populations directly influence the net current $j$ and the photovoltage $V$.
The output power generated by the quantum photocell is then given by the product of these two quantities:
\begin{eqnarray}\label{eq.14}
P_{\mathrm{out}}=j.V
\end{eqnarray}

\begin{figure}
        \centering{
        \qquad $\mathbf{(a)}$ \qquad\qquad \qquad\quad\qquad\qquad $\mathbf{(b)}$\\{
        \includegraphics[width=2.8in]{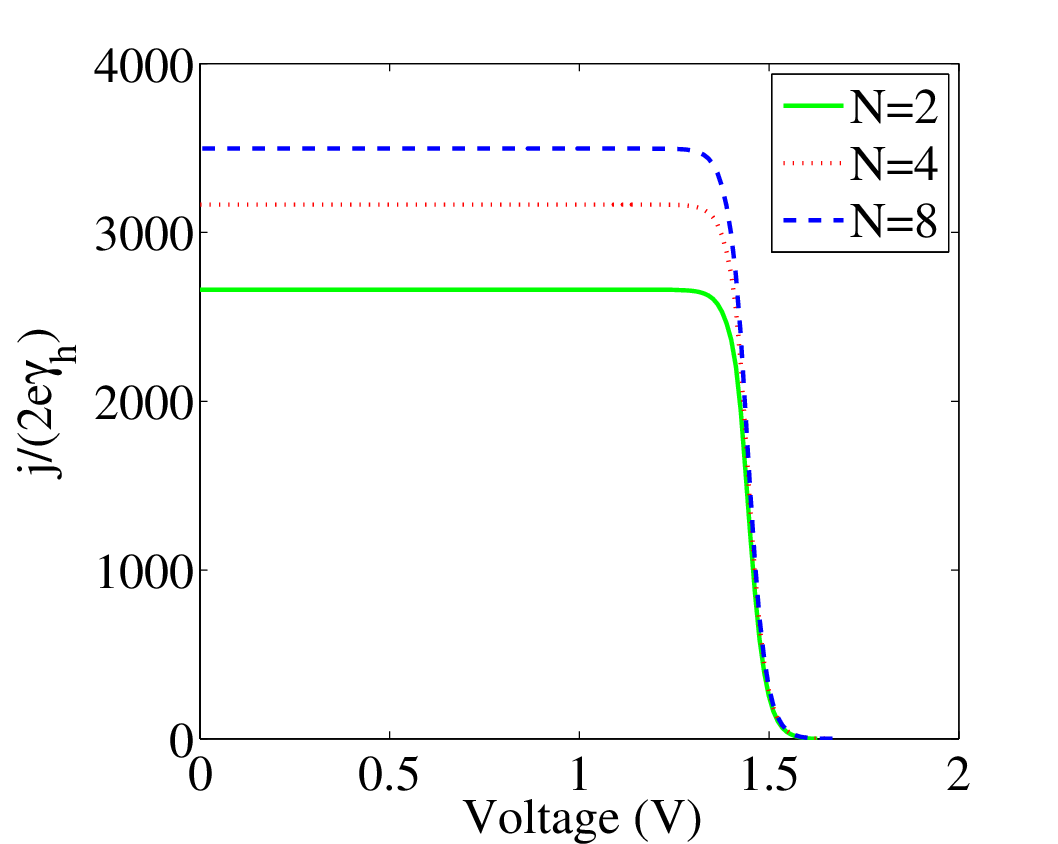}
        \label{fig:first_sub}
    }{
        \includegraphics[width=2.8in]{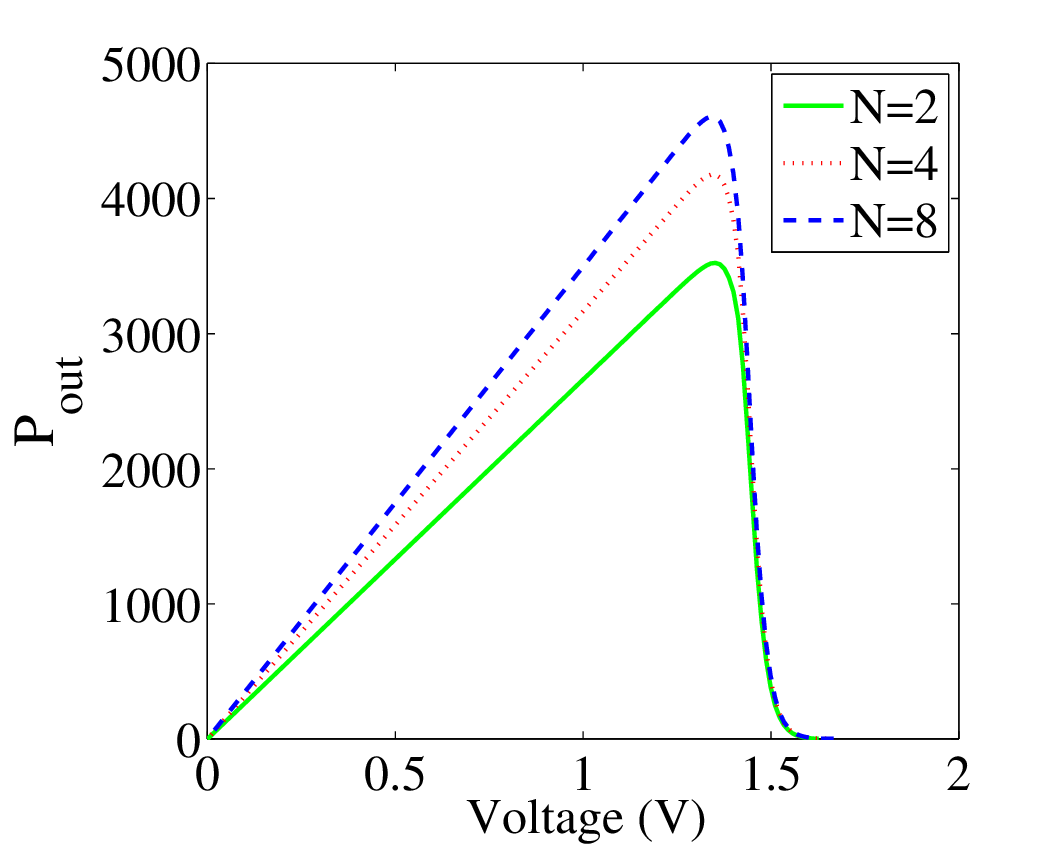}
        \label{fig:second_sub}
    }\\ \par \qquad $\mathbf{(c)}$ \qquad\qquad \qquad\quad\qquad\qquad $\mathbf{(d)}$\\{
        \includegraphics[width=2.8in]{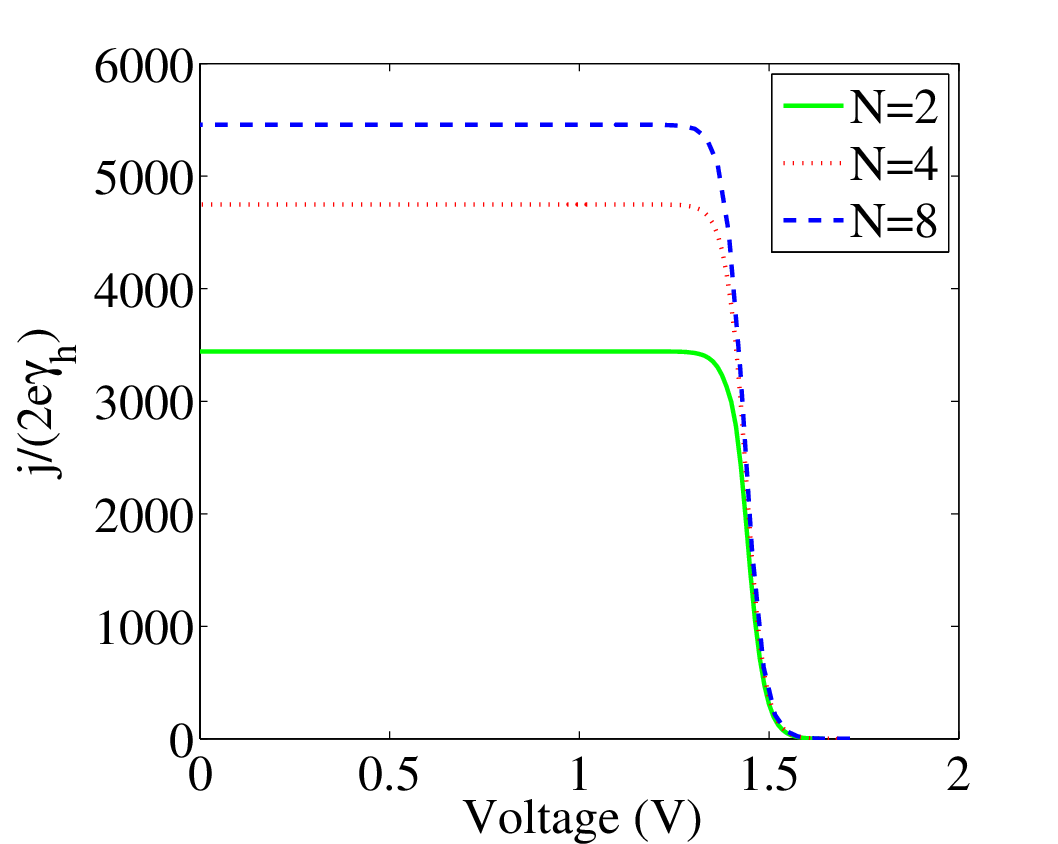}
        \label{fig:first_sub}
    }{
        \includegraphics[width=2.8in]{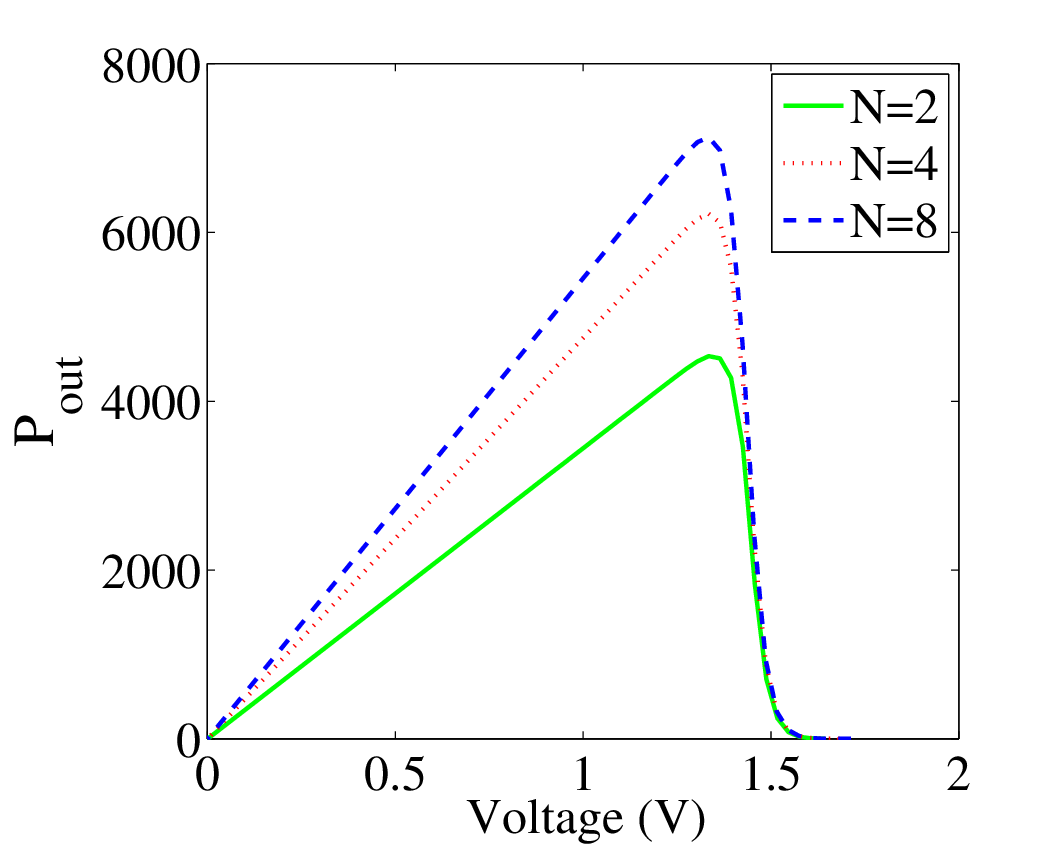}
        \label{fig:second_sub}
    }
        }
\caption{\small \bf (a,c) The steady state photocurrent $j/2e\gamma_{h}$ and (b,d) power as a function of applied voltage for respectively uncoupled and coupled donors and with three configuration of donors. (a,c) The plot illustrates the dependence of the plateau height and the onset of current suppression on the number of donor sites with higher $N$ resulting in increased current density before a sharp drop near the threshold voltage. (b,d) The slope of the initial linear region increases with $N$ then after reaching the peak drops off to zero.}
\end{figure}

Figure 2(a) illustrates the normalized electric current $j/(2e\gamma_{h})$, as a function of the output voltage $V$ for uncoupled regime with a network structure composed of $N=2$, $N=4$ and  $N=8$ electron donors. This normalization facilitates direct comparison across systems of differing complexity, while the variable $\gamma_{h}$ likely corresponds to decay rate. The extraction current was measured under ambient (room temperature) conditions $T_{c}=300 K$. The plateau height increases with the number of donors. This trend indicates that increasing the number of donor units enhances the overall exciton generation and electron transfer capacity of the network, thereby increasing the photocurrent and subsequently, the power output. According to Figure 2(a) the maximum current output $j/(2e\gamma_{h})^{Max}_{out}\approx 3496$mA of the PV device is obtained for $N=8$. Also the generation of the current output values in descending order according to multi-donors configuration is: $j/(2e\gamma_{h})_{N=8} \approx 3496 \mathrm{mA}> j/(2e\gamma_{h})_{N=4} \approx 3165 \mathrm{mA}>j/(2e\gamma_{h})_{N=2} \approx 2660 \mathrm{mA}$.
A steep decline in current is observed near $V\approx 1.67v$ for all configurations, suggesting this is the regime where the open-circuit voltage is approached. Beyond this threshold, the applied voltage opposes exciton dissociation and electron transport, leading to suppression of photocurrent. Implying that the open-circuit voltage is primarily dictated by intrinsic energy levels of the donor-acceptor system, not by donor multiplicity.
The system exhibits a sublinear $\cite{Boyd}$ enhancement of current with increasing donor number; that is, the current does not simply double when the number of donors increases from $N=2$ to $N=4$, but instead increases by a smaller factor. This sublinear scaling suggests that the contribution of additional donor units to the photocurrent becomes progressively less pronounced as the donor number increases.

Figure 2(c) illustrates the normalized electric current $j/(2e\gamma_{h})$, as a function of the output voltage $V$ for coupled regime with a network structure composed of $N=2$, $N=4$ and  $N=8$ electron donors. The current-voltage characteristics exhibit a pronounced plateau at low output voltages, followed by a sharp suppression of the photocurrent as the voltage approaches the open-circuit regime. Similar to the uncoupled regime, as the number of coupled donor sites increases, the plateau current is enhanced for $N=2$, $N=4$ and $N=8$ respectively. The current ordering in the coupled regime indicates that increasing the size of the coupled donor network significantly enhances the photocurrent in comparison with the uncoupled regime. This enhancement can be attributed to the collective nature of the donor excitations induced by dipole-dipole coupling, which modifies the excitation spectrum and facilitates more efficient redistribution and transfer of excitation energy within the donor manifold. Despite the increase in the current magnitude, the voltage at which the photocurrent is strongly suppressed remains within a relatively narrow range for the different donor configurations.
The current begins to decrease rapidly around $V\approx 1.4v$ and approaches zero at approximately $V\approx 1.67v$. This behavior indicates that the high-voltage characteristics, and particularly the open-circuit-voltage regime, are governed predominantly by the intrinsic energetic structure of the donor-acceptor system rather than by the number of donor sites alone. Importantly, increasing $N$ primarily enhances the current level while producing only a comparatively modest change in the voltage scale associated with current suppression. The results further reveal that the enhancement of the current does not scale linearly with the number of donors. In particular, doubling the number of donor sites from $N=2$ to $N=4$ produces a considerable increase in the plateau current, whereas the additional increase from $N=4$ to $N=8$ is comparatively smaller. This sublinear scaling suggests the emergence of collective effects and indicates that the benefit of adding further donor sites gradually saturates.

A direct comparison between the uncoupled and coupled donor configurations highlights the significant role of donor-donor interactions in photocurrent generation. While increasing the number of independent donors enhances the current primarily through the increased number of excitation and charge-transfer channels, the coupled configuration exhibits a markedly higher current level for comparable network sizes. Notably, the coupled $N=8$ configuration reaches a normalized plateau current of approximately $5456 \mathrm{mA}$, substantially exceeding the maximum current reported for the uncoupled $N=8$ configuration. This enhancement cannot therefore be attributed solely to the increase in donor number. Instead, it points to the additional contribution of intermolecular interactions, which generate delocalized collective excited states and modify the excitation-transfer dynamics across the donor network. Interestingly, the current enhancement with increasing $N$ in the coupled regime gradually saturates, indicating that increasing the network size beyond a certain point does not lead to a proportional increase in photocurrent. In contrast, the voltage range associated with current suppression remains relatively insensitive to donor multiplicity in both configurations, suggesting that the open-circuit behavior is primarily determined by the energetic structure of the donor-acceptor subsystem.

Figure 2(b) presents the output power $P_{out}$ as a function of voltage $V$ for three different configuration of uncoupled donors. This comparative analysis reveals how increasing the number of donor sites influences the power conversion capabilities of the device. At low voltages (approximately $V<1.32 v$) the output power increases linearly for all values of $N$. This behavior arises from the quasi-linear dependence of power on both voltage and current ($P=j.V$) and is consistent with the earlier observation that increasing $N$ enhances photocurrent generation. The slope of the initial linear region increases with $N$, indicating that multi-donor configurations enable more efficient exciton harvesting and charge transport even at low driving voltages.
For each curve, a well-defined peak is observed, marking the maximum power point, the optimal operating voltage at which the photocell delivers maximum extractable power. This maximum power point increases significantly in magnitude with increasing donor number: $P^{Max}_{N=8}\approx 4607 \mu \mathrm{W}> P^{Max}_{N=4}\approx 4177 \mu \mathrm{W}> P^{Max}_{N=2}\approx 3524 \mu \mathrm{W}$ with peaks around $V=1.35v$. The increase in peak power output with $N$ confirms that donor multiplicity enhances the collective absorption and charge transfer capacity of the network. The strong dependence of peak power on $N$ demonstrates that scaling up the donor network is an effective strategy for boosting device performance without modifying the basic materials or voltage operating range. While voltage tuning alone may not shift the upper limits of output voltage, architectural design (increasing $N$) offers a practical pathway to increase the total extractable power.

Figure 2(d) presents the output power $P_{out}$ as a function of voltage $V$ for three different configuration of coupled donors.
For all donor numbers, the output power initially increases approximately linearly with voltage, while the increasing slope with $N$ reflects the enhanced current-generating capability of the enlarged donor network. The maximum output power increases from approximately $4533 \mu \mathrm{W}$ for $N=2$ to $6221 \mu \mathrm{W}$ and $7128 \mu \mathrm{W}$ for $N=4$ and $N=8$, respectively, whereas the corresponding maximum-power voltages remain within a relatively narrow range. Importantly, this enhancement should not be interpreted solely as a consequence of increasing the number of donor sites. The dipole-dipole interaction reorganizes the donor excitation manifold and generates collective eigenstates with modified energies and delocalized excitation character. Consequently, the excitation can be distributed coherently over multiple donor molecules rather than remaining localized on individual sites. Such coherent superposition and delocalization provide additional collective pathways for excitation redistribution and charge transfer toward the acceptor, thereby allowing the enlarged donor ensemble to contribute more effectively to power generation. The progressively higher power obtained with increasing $N$ therefore reflects the combined contribution of donor multiplicity and intermolecular quantum effects, highlighting that the performance scaling of the coupled photocell arises from a genuinely collective response of the donor network rather than from a simple accumulation of independent donor contributions.

A direct comparison of the output-power characteristics of the uncoupled and coupled donor configurations reveals a clear advantage of the latter architecture. In the uncoupled system, increasing the number of donors enhances the output power primarily by increasing the number of available excitation and charge-transfer channels, resulting in a higher initial power slope and an increased maximum power. However, the enhancement remains essentially a consequence of the additive contribution of independently evolving donor sites. In contrast, the coupled configuration exhibits a higher output power for comparable donor numbers, indicating that the performance enhancement cannot be explained solely by donor multiplicity. The dipole-dipole interaction introduces a qualitatively different quantum regime by reorganizing the donor excitation manifold into collective eigenstates with modified energies and delocalized character. In particular, the resulting coherent superposition of donor excitations and their spatial delocalization enable the donor ensemble to participate collectively in excitation redistribution and charge-transfer processes. Thus, the coupled architecture provides an additional degree of control over the energy-transfer pathways that is absent in the uncoupled configuration.

In this paper, we define the relative power enhancement as the difference between the output powers of the coupled and uncoupled regimes, normalized by the output power of the uncoupled configuration and expressed as a percentage. This quantity characterizes the enhancement in the output power arising from the intermolecular coupling and the associated quantum effects, and is given by

\begin{eqnarray}\label{Equation4}
PE=(\frac{P^{\mathrm{coupled}}_{\mathrm{out}}-P^{\mathrm{uocoupled}}_{\mathrm{out}}}{P^{\mathrm{uocoupled}}_{out}})\times100,
\end{eqnarray}
where $P^{\mathrm{coupled}}$ and $P^{\mathrm{uocoupled}}$ being the power output in the coupled and uncoupled case, respectively. It is important to note that, in the absence of intermolecular donor-donor coupling, the dynamics of the system reduce to those of a classical cyclic heat engine, as the donor molecules evolve independently. In this limit, the Lindblad equation governing the level populations contains no off-diagonal contributions, and the resulting population dynamics can consequently be described by a Pauli master equation, providing a classical description of the cyclic engine. In contrast, when the dipole-dipole interaction is finite, quantum correlations between the donor excitations are incorporated into the system dynamics through the emergence of off-diagonal density-matrix elements. The corresponding evolution therefore captures the quantum-coherent regime of the cyclic engine, in which the donor molecules are coupled and their collective excitation dynamics can contribute to the energy-transfer and power-generation processes. A key finding of this study is the substantial enhancement in photovoltaic efficiency induced by intermolecular dipole-dipole coupling. Quantitatively, the coupled configuration yields efficiency enhancements of $28.63\%$, $48.93\%$, and $54.72\%$ for $N=2$, $N=4$, and $N=8$ donor molecules, respectively, relative to their corresponding uncoupled configurations. Notably, the enhancement increases with donor multiplicity, demonstrating that the incorporation of coherent interactions can provide a significant contribution to the photovoltaic performance of the multi-donor quantum photocell.

Figure 3 illustrates the normalized photocurrent  $j/(2e\gamma_{h})$ as a function of the number of donor sites $N$ (ranging from 2 to 10) for both the uncoupled and coupled regimes where the output voltage remains approximately constant $V=1.35v$ within the quasi-linear domain in multi-donor quantum photocell model. This voltage corresponds to the operational regime of the photocell prior to the onset of saturation and current quenching effects near the open-circuit voltage threshold. Both curves exhibit a monotonically increasing, nonlinear trend, indicating that the photocurrent enhances progressively as additional donor units are introduced into the quantum photocell architecture. In the coupled case, where coherent interactions among the $N$ donor excited states are retained, the current rises sharply from $N=2$ to $N=10$. In contrast, the uncoupled curve increases only modestly over the same range.
The persistent vertical separation between the two curves demonstrates that quantum coherence among the excited donor levels systematically enhances charge-transfer efficiency. The observed enhancement originates from the formation of delocalized collective states. Constructive interference within these states accelerates the effective donor-acceptor transfer rate while suppressing competing loss channels.
As $N$ increases, the collective manifold becomes richer, allowing a larger fraction of the excitation energy to be funneled into the photocurrent pathway; this is reflected in the steeper initial slope and higher saturation value of the coupled curve. These results establish that quantum coherence and collective coupling remain advantageous even when the donor number is scaled well beyond the minimal two-site configuration. The progressive current gain confirms that the multi-donor architecture can be engineered to exploit collective quantum effects for improved photovoltaic performance.

\begin{figure}
\centering $\mathbf{(a)}$\\{
\includegraphics[width=4.4in]{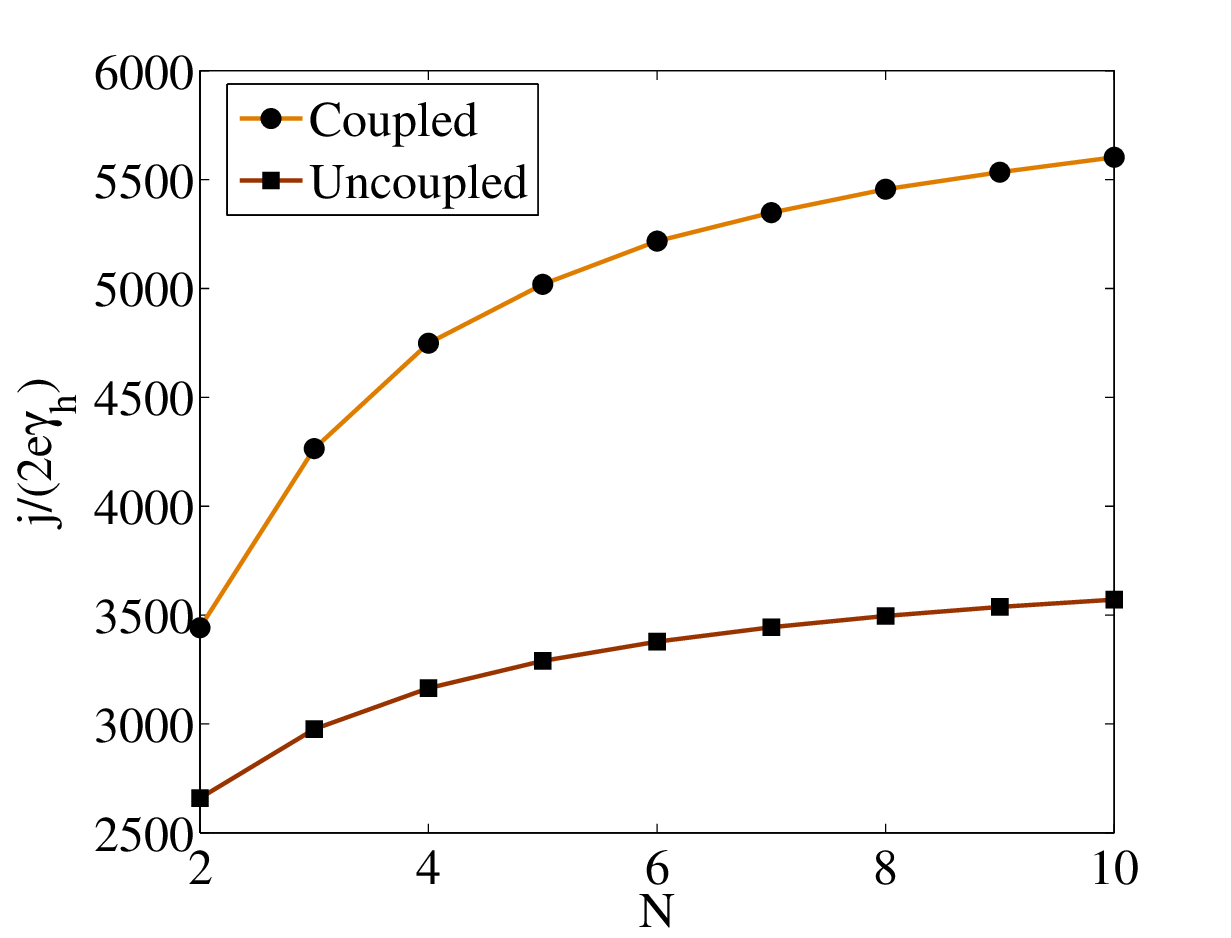}
\label{fig:first_sub}
}
\caption{\small \bf Normalized photocurrent  versus donor number $N$ in the constant-voltage regime $(V=1.35v)$ shows a sublinear increase at low $N$, followed by saturation at higher $N$. This reflects enhanced exciton dynamics and charge separation at low $N$, and performance limits due to acceptor saturation and extraction bottlenecks at high $N$.}
\end{figure}

\section{CONCLUSIONS}
The present study establishes, for the first time, a comprehensive quantum-mechanical treatment of a multi-donor photocell comprising an arbitrary number $N$ of electron donors coupled to a single acceptor. By explicitly retaining inter-donor coherence and the resulting collective excited states, we demonstrate that constructive quantum interference opens optimal energy-transfer pathways that are inaccessible in the incoherent limit. Quantitative analysis of the steady-state photocurrent and output power reveals that the coherently coupled network systematically outperforms its uncoupled counterpart for every examined $N$. The enhancement grows with donor multiplicity, saturating only after a pronounced rise that reflects the progressive delocalization of excitation across the collective manifold. Critically, this performance gain is achieved without degradation of the open-circuit voltage, confirming that the underlying energy-level architecture remains robust under increasing system size. These results close a long-standing computational gap: previous investigations were effectively limited to $N\leq 3$ by the rapid growth of Hilbert-space dimension and the technical demands of treating many-body coherence. The framework developed here removes that barrier, furnishing a scalable, fully quantum description of the $N$-donor photocell. In doing so, it identifies collective coherence not merely as a perturbative correction but as a decisive design principle for next-generation quantum photovoltaic architectures.

\section*{Author contributions}
B.M.J.N.: Formal analysis, Investigation, Methodology, Validation, Writing-original draft.
E.F.: Investigation, Project administration, Resources, Validation.
B.A.: Conceptualization, Formal analysis, Investigation, Methodology, Software, Supervision, Validation, Visualization, Writing-review and editing.

\section*{Data availability}
The data of the present study are available from the corresponding author upon a reasonable request.

\section*{Conflict of interest}
The authors declare that they have no conflict of interest.

\end{document}